\documentclass[aps,prl,twocolumn,superscriptaddress,amsmath,amssymb]{revtex4}
\usepackage{graphicx,epsfig}
\usepackage{color}
\usepackage{bm}
\usepackage[normalem]{ulem}

\newcommand{\be}{\begin{equation}}
\newcommand{\bea}{\begin{eqnarray}}
\newcommand{\eea}{\end{eqnarray}}
\newcommand{\ee}{\end{equation}}

\begin{document}

\title{Deep inner-shell multiphoton ionization by intense x-ray free-electron laser pulses}

\author{H.~Fukuzawa}
\affiliation{Institute of Multidisciplinary Research for Advanced
Materials, Tohoku University, Sendai 980-8577, Japan}
\affiliation{RIKEN SPring-8 Center, Kouto 1-1-1, Sayo, Hyogo 679-5148, Japan}
\author{S.-K.~Son}
\affiliation{Center for Free-Electron Laser Science (CFEL), DESY, 
22607 Hamburg, Germany}
\author{K.~Motomura}
\author{S.~Mondal}
\affiliation{Institute of Multidisciplinary Research for Advanced
Materials, Tohoku University, Sendai 980-8577, Japan}
\author{K.~Nagaya}
\affiliation{RIKEN SPring-8 Center, Kouto 1-1-1, Sayo, Hyogo 679-5148, Japan}
\affiliation{Department of Physics, Kyoto University, Kyoto 606-8502, Japan}
\author{S.~Wada}
\affiliation{RIKEN SPring-8 Center, Kouto 1-1-1, Sayo, Hyogo 679-5148, Japan}
\affiliation{Department of Physical Science, Hiroshima University, Higashi-Hiroshima 739-8256, Japan}
\author{X.-J.~Liu}
\affiliation{Synchrotron SOLEIL, L'Orme des Merisiers, Saint-Aubin, BP 48, FR-91192 Gif-sur-Yvette Cedex, France}
\author{R.~Feifel}
\affiliation{Department of Physics and Astronomy, Uppsala University,
P.O. Box 516, SE-751 20 Uppsala, Sweden}
\author{T.~Tachibana}
\author{Y.~Ito}
\author{M.~Kimura}
\affiliation{Institute of Multidisciplinary Research for Advanced
Materials, Tohoku University, Sendai 980-8577, Japan}
\author{T.~Sakai}
\author{K.~Matsunami}
\affiliation{Department of Physics, Kyoto University, Kyoto 606-8502, Japan}
\author{H.~Hayashita}
\author{J.~Kajikawa}
\affiliation{Department of Physical Science, Hiroshima University, Higashi-Hiroshima 739-8256, Japan}
\author{P.~Johnsson}
\affiliation{Department of Physics, Lund University, P.O. Box 118, 22100 Lund, Sweden}
\author{M.~Siano}
\affiliation{Blackett Laboratory, Imperial College London, London, United Kingdom}
\author{E.~Kukk}
\affiliation{Department of Physics and Astronomy, University of Turku, 20014, Finland}
\author{B.~Rudek}
\author{B.~Erk}
\affiliation{Max-Planck Advanced Study Group at CFEL, 22607 Hamburg, Germany}
\affiliation{Max-Planck-Insitut f{\"u}r Kernphysik, 69117 Heidelberg, Germany}
\author{L.~Foucar}
\affiliation{Max-Planck Advanced Study Group at CFEL, 22607 Hamburg, Germany}
\affiliation{Max-Planck-Insitut f{\"u}r medizinische Forschung, 69120 Heidelberg, Germany}
\author{E.~Robert}
\author{C.~Miron}
\affiliation{Synchrotron SOLEIL, L'Orme des Merisiers, Saint-Aubin, BP 48, FR-91192 Gif-sur-Yvette Cedex, France}
\author{K.~Tono}
\affiliation{Japan Synchrotron Radiation Research Institute (JASRI), 
Kouto 1-1-1, Sayo, Hyogo 679-5198, Japan}
\author{Y.~Inubushi}
\author{T.~Hatsui}
\author{M.~Yabashi}
\affiliation{RIKEN SPring-8 Center, Kouto 1-1-1, Sayo, Hyogo 679-5148, Japan}
\author{M.~Yao}
\affiliation{Department of Physics, Kyoto University, Kyoto 606-8502, Japan}
\author{R.~Santra}
\email[]{robin.santra@cfel.de}
\affiliation{Center for Free-Electron Laser Science (CFEL), DESY, 
22607 Hamburg, Germany}
\affiliation{Department of Physics, University of Hamburg, 20355 Hamburg, Germany}
\author{K.~Ueda}
\email[]{ueda@tagen.tohoku.ac.jp}
\affiliation{Institute of Multidisciplinary Research for Advanced
Materials, Tohoku University, Sendai 980-8577, Japan}
\affiliation{RIKEN SPring-8 Center, Kouto 1-1-1, Sayo, Hyogo 679-5148, Japan}

\date{\today}

\begin{abstract}
We have investigated multiphoton multiple ionization dynamics of argon and xenon atoms using a new x-ray free electron laser (XFEL) facility, SPring-8 {\AA}ngstrom Compact free electron LAser (SACLA) in Japan, and identified that Xe$^{n+}$ with $n$ up to 26 are produced predominantly via four-photon absorption as well as Ar$^{n+}$ with $n$ up to 10 are produced via two-photon absorption at a photon energy of 5.5~keV.
The absolute fluence of the XFEL pulse, needed for comparison between theory and experiment, has been determined using two-photon processes in the argon atom with the help of benchmark {\it ab initio} calculations.
Our experimental results, in combination with a newly developed theoretical model for heavy atoms, demonstrate the occurrence of  multiphoton absorption involving deep inner shells.  
\end{abstract}


\maketitle

Multiphoton processes in the optical regime are well-known phenomena investigated for decades. 
The advent of extreme ultraviolet (EUV)~\cite{FLASH,SCSS} and x-ray~\cite{LCLS} free-electron lasers (FELs), with femtosecond pulse widths, has led to renewed interest in multiphoton processes in the EUV to x-ray spectral region.  
See, for example, some of the recent works on atoms and small molecules carried out at FLASH~\cite{FLASH} in Germany~\cite{Richter2009PRL,Kurka2010NJP,Rouzee2011PRA} and at the SCSS test accelerator~\cite{SCSS} in Japan~\cite{Motomura2009JPB,Yamada2010JCP,Gryzlova2011PRA,Sato2011JPB,Hishikawa2011PRL}, as well as at LCLS~\cite{LCLS} in USA~\cite{Young2010Nature,Hoener2010PRL,Fang2010PRL,Cryan2010PRL,Berrah2011PNAS,Doumy2011PRL,Salen2012PRL,Rudek2012submitted}.
The motivation for these studies has been not only to reveal these pathways of the multiphoton multiple ionization newly opened by these light sources (see, e.g., \cite{Richter2009PRL,Gryzlova2011PRA,Hishikawa2011PRL,Young2010Nature,Doumy2011PRL,Rohringer2007PRA,Makris2009PRL,Lambropoulos2011PRA,Son2011PRA,Son2012PRA}) but also to employ these processes as the basis for a new type of spectroscopy for chemical analysis (see, e.g., \cite{Berrah2011PNAS,Salen2012PRL,Santra2009PRL}).
The relevance of multiphoton multiple ionization processes for electronic radiation damage in materials has also been noted~\cite{Hoener2010PRL,Son2011PRL}.
Electronic radiation damage due to multiphoton processes is a crucial issue for x-ray imaging using an XFEL. 
So far, however, multiphoton experiments have been limited to the photon energy range up to 2~keV, i.e., the upper photon energy limit of the atomic, molecular, and optical physics beam line at LCLS.

In March 2012, a new XFEL facility, the SPring-8 {\AA}ngstrom Compact free electron LAser (SACLA)~\cite{Ishikawa2012NaturePhoton}, started user operation in Japan.
Using this new facility, we have investigated multiphoton multiple ionization dynamics of argon and xenon atoms in intense hard x-ray pulses.
The Ar $K$-shell thresholds are around 3~keV, and the Xe $L$-shell thresholds are around 5~keV.
The intense x-ray pulses from SACLA, with a photon energy above those deep inner-shell thresholds, induce complex multiple ionization dynamics characterized by the absorption of several photons.
We identify that for a single Xe atom, absorption of 4 photons of 5.5~keV induces emission of up to 26 electrons.
As a consequence of intraatomic electron--electron interactions, each photon causes the ejection of more than 6 electrons on average.

From the pioneering work on the light neon atom, we learned that x-ray multiphoton multiple ionization is well characterized by a sequence of one-photon ionization accompanied by decay processes~\cite{Rohringer2007PRA,Young2010Nature}.
The complexity increases for heavy atoms, where one-photon ionization of a deep inner shell initiates a decay cascade, i.e., a series of decay steps leading to the emission of several electrons~\cite{Carlson1966PR}.
The extremely large number of x-ray photons in an XFEL pulse is responsible for triggering further photoionization from deep inner shells during and after a decay cascade.
Thus sequential multiphoton absorption involving deep inner shells becomes very complicated.
As a consequence of its sequential nature, x-ray multiphoton absorption depends primarily on the fluence, which is the number of photons (or the pulse energy) per unit area, rather than the (peak) intensity~\cite{Doumy2011PRL,Rudek2012submitted}.
The minimum fluence required for multiphoton absorption is estimated as follows.
For example, the photoionization cross section of neutral Xe at 5.5~keV is $\sim$0.166~Mb, so the fluence to saturate one-photon 
absorption is $\sim$6$\times$10$^{10}$~photons/$\mu$m$^2$, which corresponds to $\sim$50~$\mu$J/$\mu$m$^2$ for 5.5-keV photons. 
If the fluence of an x-ray pulse is close to or higher than this fluence, one expects that more than one photon will be absorbed by a xenon atom within one pulse.

The experiment has been carried out at the experimental hutch 3 (EH3) of the beam line 3 (BL3) of the SACLA in Japan~\cite{Ishikawa2012NaturePhoton}.
The photon energy was set at 5.5~keV. 
The photon band width was $\sim$60~eV (FWHM). 
The repetition rate of the XFEL pulses was 10~Hz. 
The pulse width has not been measured, but was estimated to be in the range between 10 and 30~fs (FWHM).  
Shot-by-shot pulse energy was measured by the beam-position monitor~\cite{Tono2011RSI} located upstream of the beam line described below.
The monitor was calibrated by a calorimeter~\cite{Kato2012APL} so that output signals from the monitor could be transformed to 
the absolute value of the pulse energy.
The measured values were 239~$\mu$J per pulse on average.

\begin{figure}
\begin{center}
\includegraphics[width=8cm]{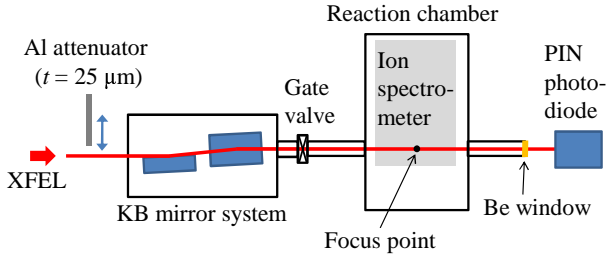}
\end{center}
\caption{(Color online) Experimental configuration.}
\label{Fig:Config}
\end{figure}

Figure~\ref{Fig:Config} shows the experimental configuration. 
A Kirkpatrick-Baez (KB) mirror system is permanently installed at EH3~\cite{RIKENweb}. 
The focal length (work distance) is $\sim$1.3~m.
The XFEL beam is focused by the KB mirror system to a focal size of $\sim$1~$\mu$m (FWHM) in diameter. 
The Rayleigh length is $\sim$8~mm. 
The sample gas (argon or xenon) was introduced as a pulsed supersonic gas jet~\cite{Nagaya2010JESRP} to the focus point of the XFEL pulses, in the ultrahigh-vacuum reaction chamber.
The gas beam at the reaction point was estimated to be $\sim$2~mm (FWHM) in diameter. 
Thus, the source volume of the ions was roughly a cylindrical shape of $\sim$1~$\mu$m in diameter and $\sim$2~mm along the XFEL beam.
After crossing the gas jet at right angles, the XFEL beam came out of the vacuum chamber via a beryllium window.
The relative x-ray pulse energy was measured shot-by-shot by a p-intrinsic-n (PIN) photodiode, after the pulse energy was reduced by aluminum sheets of 0.2--0.5~mm thickness so that linear response of the photodiode was assured.
The shot-by-shot pulse energy fluctuation was $\pm$25\% (50\% FWHM).  

\begin{figure}
\begin{center}
\includegraphics[width=8cm]{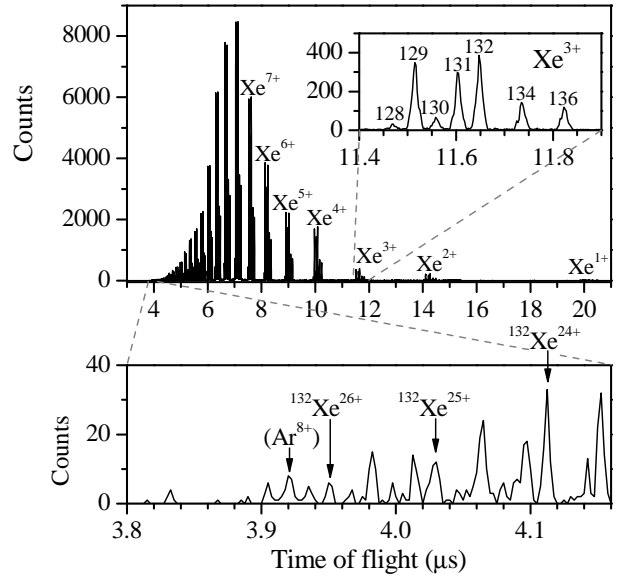}
\end{center}
\caption{The ion time of flight spectrum of Xe recorded at the photon energy of 5.5~keV at full XFEL pulse energy.}
\label{Fig:TOF_Xe}
\end{figure}

Ions produced in the source volume described above were extracted towards the ion time-of-flight (TOF) spectrometer~\cite{Liu2009RSI} equipped with microchannel plates (MCP) and a delay-line anode (Roentdek HEX80)~\cite{Jagutzki2002IEEE}.
Signals from the MCP and the delay-line anode were fed to an 8-channel digitizer. 
The wave forms recorded by the digitizer were analyzed by a software discriminator~\cite{Motomura2009NIM} and the arrival time and the arrival position of each ion were determined.
The voltages on the spectrometer are tuned for the best mass resolution. 
The simulation of ion trajectories shows that the TOF slightly depends on the ion arrival position on the detector, as a consequence of different departure positions perpendicular to spectrometer axis. 
A compensation on the measured TOF is introduced to improve the mass resolution further. 
Figure~\ref{Fig:TOF_Xe} depicts the TOF spectrum for xenon ions after compensation of the TOF.    
We can clearly see ions with a charge state up to +26 and well-resolved isotopes at different charge states.

Let us now describe our theoretical approach to x-ray multiphoton multiple ionization dynamics.
We employ the \textsc{xatom} toolkit~\cite{Son2011PRA}, based on the rate equation approach~\cite{Rohringer2007PRA} and the Hartree--Fock--Slater method.  
For Ar, we use the approach described in Ref.~\cite{Son2011PRA}, adding shakeoff processes within the sudden approximation~\cite{Kochur1994JPB}.
For Xe, even this straightforward sequential ionization model becomes tremendously challenging, because (a) there are too many electronic configurations involving multiple holes and too many atomic data (photoionization cross sections, Auger and Coster--Kronig rates, and fluorescence rates), and (b) the matrix size for the set of coupled rate equations is too large to be solved directly.
We have addressed the latter by introducing a Monte Carlo approach in Refs.~\cite{Son2012PRA,Rudek2012submitted} where for Xe $M$-shell ionization dynamics $\sim$$10^6$ coupled rate equations were solved and $\sim$4$\times$10$^7$ atomic data were pre-calculated.
For Xe $L$-shell ionization dynamics considered here, however, the complexity is further increased by a factor of 21, thus it becomes formidable to pre-calculate all the atomic data required for the rate equations.
Therefore, we extend \textsc{xatom} by applying the Monte Carlo procedure for both calculating atomic data and searching probable ionization pathways. 
When a certain configuration is visited during an ionization pathway chosen by the Monte Carlo sampling, atomic data are calculated for the corresponding configuration.
In this way, atomic data are computed only when they are required. 
This extension enables us to treat ionization dynamics of heavy atoms with no limit of configurational space.
Saving in computational time is dramatic.
For example, one calculation of Xe at 5.5~keV and 190~$\mu$J/$\mu$m$^2$ takes 2 days using one CPU on the laboratory workstation, which would take more than 3 years if all atomic data had to be calculated.

In the present calculations, the photon energy is fixed at 5.5~keV. 
The pulse shape is assumed to be a Gaussian of 30 fs (FWHM). 
In the regime of sequential multiphoton absorption, the results are not sensitive to the pulse shape or spikiness of individual pulses~\cite{Rohringer2007PRA}.
We also did the calculation for 10~fs (FWHM) at a single XFEL fluence and confirmed that the results are in good agreement. 
All calculated results presented here are obtained by 3-dimensional integration~\cite{Rudek2012submitted} over the interaction volume of 1~$\mu$m $\times$ 1~$\mu$m $\times$ 2~mm (all lengths are given in FWHM), according to the instrumental configuration.

In order to compare the theoretical results with the experimental results, we need the peak fluence of the XFEL pulse employed for our experiment.
The fluence at the center of the x-ray beam focus defines the peak fluence $F_{\rm {peak}}$, which is given by
\[
F_{\rm {peak}} = \frac{4 \ln 2}{\pi} \times \frac{E}{A} \times T,
\]
where $E$ is the nominal pulse energy from the monitor, $T$ is the transmission(\%), $A$ is the focal area given in FWHM$\times$FWHM, and the coefficient of $4 \ln 2 / \pi$ comes from the assumed Gaussian focal shape.
In order to estimate the peak fluence of the XFEL pulse, we have measured the charge state distribution of Ar atoms and compared it with our benchmark calculations. 
At 5.5~keV, the $K$-shell electrons of Ar can be ionized.
The number of all possible multiple-hole configurations, or equivalently the number of coupled rate equations is 1323, which is easily solvable without the Monte Carlo method.

\begin{figure}
\begin{center}
\includegraphics[width=8cm]{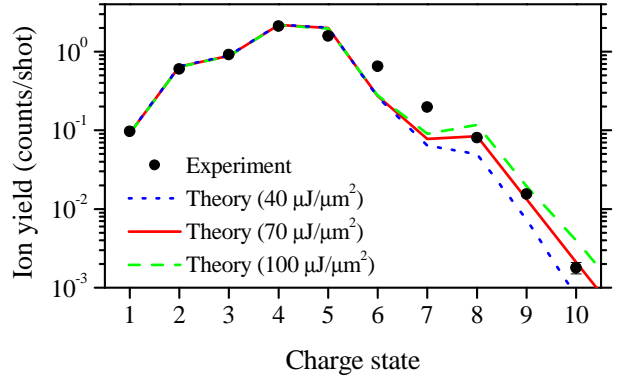}
\end{center}
\caption{(Color online) Experimental and theoretical charge state distributions of Ar at the photon energy of 5.5~keV.}
\label{Fig:CSD_Ar}
\end{figure}

The experimental and theoretical charge state distributions of Ar are shown in Fig.~\ref{Fig:CSD_Ar}. 
Theoretical charge state distributions are scaled so that the sum of the individual charge state yields is equal to the total ion yield in the experiment.
We find that the theoretical distributions up to the charge state of +6 do not vary when varying the XFEL fluence, illustrating that these ions are produced by single photon absorption. 
In general, they are in good agreement with experiment, except that in theory the Ar$^{5+}$ is slightly overestimated while Ar$^{6+}$ and Ar$^{7+}$ are underestimated.
Further, we find that both experimental and theoretical yields of Ar$^{8+}$ and Ar$^{9+}$ depend quadratically on the XFEL fluence. 
We thus can estimate relative contributions from two- and one-photon processes using the ratio of $R \equiv (Y_{8+} + Y_{9+})/(Y_{3+} + Y_{4+})$, where $Y_{n+}$ is the yield of Ar$^{n+}$.
Fitting the theoretical value of $R$ to the experimental one by the least squares method, we have obtained 70~$\mu$J/$\mu$m$^2$ for the peak fluence.
The uncertainty of this estimate mainly stems from our benchmark theoretical calculations. 
We expect that this uncertainty is $\sim$$\pm$10\%.
The resulting value of 70$\pm$7~$\mu$J/$\mu$m$^2$ is consistent with the empirical knowledge from other experiments, that the transmission of the KB mirror system is $\sim$50\% and that the focal area is $\sim$2~$\mu$m$^2$, though these numbers may include uncertainties of a factor of two.

\begin{figure}
\begin{center}
\includegraphics[width=8cm]{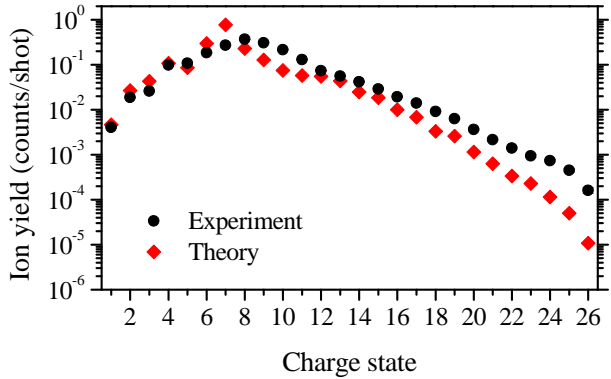}
\end{center}
\caption{(Color online) Experimental and theoretical charge state distributions of Xe at the photon energy of 5.5~keV with the peak fluence of 70~$\mu$J/$\mu$m$^2$.}
\label{Fig:CSD_Xe}
\end{figure}

Figure~\ref{Fig:CSD_Xe} depicts the charge state distribution of Xe at the peak fluence of 70~$\mu$J/$\mu$m$^2$, which is determined via the calibration using Ar.
The charge state distribution varies as the peak fluence varies (not shown here).
We also compare the theoretical charge state distributions at this peak fluence with experiment. 
The photon energy of 5.5~keV is above the $L$-shell threshold for charge states up to +23 according to our calculations.
There is no signature of resonance-enabled ionization enhancement~\cite{Rudek2012submitted} because the fluence is not enough to form high charge states that initiate resonance excitation.
The discrepancy between theory and experiment may be attributed to the non-relativistic treatment and lack of shakeoff in the current theoretical model.  
The shakeoff process can further ionize valence electrons after photoionization, and some decay channels might be absent without relativity~\cite{Son2012PRA}. 
Inclusion of both relativity and shakeoff tends to produce higher charge states. 
Thus, in the current model, the formation of high charge states is somewhat suppressed in comparison with the experimental results. 
In spite of these limitations of the current model, the experimental and theoretical results are in reasonable agreement, at least semiquantitatively. 

\begin{figure}
\begin{center}
\includegraphics[width=8cm]{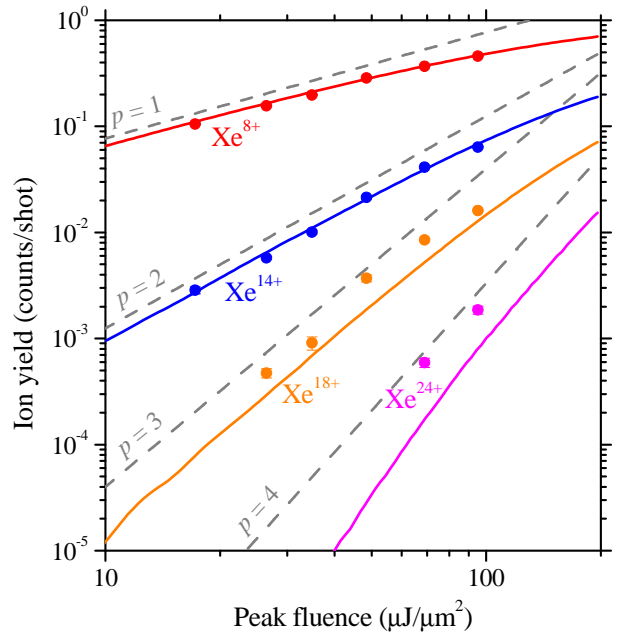}
\end{center}
\caption{(Color online) XFEL fluence dependence of the ion yields for Xe$^{n+}$ ($n$=8, 14, 18, and 24).
Closed circles with the error bars depict the experimental results and solid lines depict the theoretical results.  
Lines with slope $p$ = 1, 2, 3, and 4 are also shown as broken lines to guide the eye.
The uncertainty of the peak fluence is expected to be $\sim$$\pm$10\%.} 
\label{Fig:PowDep_Xe}
\end{figure}

To obtain the fluence dependence of the yields for the xenon ions at different charge states in a pulse energy range wider than the shot-by-shot pulse energy fluctuation, we recorded the spectra not only at the full XFEL pulse energy but also at a pulse energy attenuated by an aluminum foil of 25~$\mu$m thickness located upstream of the KB mirror system (see Fig.~\ref{Fig:Config}). 
The attenuated pulse energy was 38\% of the full pulse energy according to the reading of the PIN photodiode.
In Fig.~\ref{Fig:PowDep_Xe}, the ion yields for Xe$^{n+}$ ($n$=8, 14, 18, and 24) are plotted as a function of the peak fluence.
To obtain several data points with respect to the peak fluence, we first merged the results measured at two different pulse energies, and then re-binned the data. 
Straight lines with a slope of 1, 2, 3, and 4 are also shown as a guide to the eye.
The fluence given on the horizontal axis is on the absolute scale determined by the Ar calibration.
All theoretical ion yields are scaled by a single factor such that the yield of Xe$^{8+}$ at 70~$\mu$J/$\mu$m$^2$ is matched with that in the experiment.
The yield of Xe$^{8+}$ exhibits a slope of less than one, illustrating that this ion is produced by single-photon absorption and that the single-photon absorption process is close to saturation.
The yield of Xe$^{14+}$ exhibits a slope close to two,
the yield of Xe$^{18+}$ exhibits a slope close to three,
and the yield of Xe$^{24+}$ exhibits a slope close to four.
These slopes directly suggest that these ions are predominantly produced by two-, three-, and four-photon absorption, respectively. 
Although theory reproduces the charge state distributions only semiquantitatively, it reproduces very well the fluence dependence of the individual charge states. 
This agreement confirms the number of absorbed photons yielding the individual charge states and thus fully confirms that $L$-shell multiphoton processes of the xenon atom take place in the x-ray regime at 5.5~keV.

It is worth noting the relevance of the present work to the emerging area of femtosecond crystallography with XFELs~\cite{Chapman2011Nature,Boutet2012Science}.
Electronic radiation damage, especially to heavy atoms, is inevitable during femtosecond XFEL pulses, as seen in the present work on the xenon atom.
The high-intensity version of multiwavelength anomalous diffraction~\cite{Son2011PRL} beneficially exploits multiphoton multiple ionization dynamics of heavy atoms embedded into macromolecules and provides a new path to determine macromolecular structure using XFELs.
Therefore, a comprehensive picture of multiphoton multiple ionization dynamics is crucial to success in femtosecond x-ray imaging with XFELs.

In summary, we have investigated multiphoton multiple ionization dynamics of the xenon atom using the new XFEL facility SACLA as well as a newly developed theoretical model, and identified that Xe$^{n+}$ with $n \ge 24$ are produced predominantly via four-photon absorption, demonstrating the occurrence of multiphoton absorption of a single atom in the x-ray regime above 2~keV for the first time. 
The absolute fluence of the XFEL pulses has been determined using two-photon processes in argon atoms with the help of benchmark {\it ab initio} calculations.

The experiments were performed at SACLA with the approval of 
JASRI and the program review committee (No.~2012A8036).
This study was supported by the X-ray Free Electron Laser Utilization 
Research Project and of the Ministry of Education, Culture, Sports, 
Science and Technology of Japan (MEXT), by the Japan Society for 
the Promotion of Science (JSPS), and by the IMRAM project.

\end{document}